# Electron and proton irradiation effect on the minority carrier lifetime in SiC passivated p-doped Ge wafers for space photovoltaics

Charlotte Weiss[1], Seonyong Park[2], Jérémie Lefèvre[2], Bruno Boizot[2], Christian Mohr[1], Olivier Cavani[2], Sandrine Picard[3], Rufi Kurstjens[4], Tim Niewelt[1], Stefan Janz[1]

[1]*Fraunhofer Institute for Solar Energy Systems, 79110 Freiburg, Germany*
[2]*Laboratoire des Solides Irradiés, CNRS-UMR 7642, CEA-DRF-IRAMIS, Ecole Polytechnique, Université Paris-Saclay, Palaiseau Cedex, 91120, France*
[3]*CSNSM, Université Paris-Sud, CNRS/IN2P3, Université Paris-Saclay, 91405, Orsay, France*
[4]*Umicore Electro - optic Materials, 2250, Olen, Belgium.*

## Abstract

We report on the effect of electron and proton irradiation on effective minority carrier lifetimes ($\tau_{eff}$) in p-type Ge wafers. Minority carrier lifetimes are assessed using the microwave-detected photoconductance decay (µW-PCD) method. We examine the dependence of $\tau_{eff}$ on the p-type doping level and on electron and proton radiation fluences at 1 MeV. The measured $\tau_{eff}$ before and after irradiation are used to estimate the minority carriers' diffusion lengths, which is an important parameter for solar cell operation. We observe $\tau_{eff}$ ranging from ≈ 50 to 230 µs for Ge doping levels between 1x10$^{17}$ and 1x10$^{16}$ at.cm$^{-3}$, corresponding to diffusion lengths of ≈ 500-1400 µm. A separation of $\tau_{eff}$ in Ge bulk lifetime and surface recombination velocity is conducted by irradiating Ge lifetime samples of different thicknesses. The possible radiation-induced defects are discussed on the basis of literature.

## 1 Introduction

The research on solar cells for space applications is increasingly focusing on higher efficiencies and thinner cells in order to increase the power-to-weight (W/g) ratio [1]. Today's state-of-the-art technology for space is a triple junction (TJ) GaInP/GaInAs/Ge cell, which is current-limited by the topcell, whereas the Ge subcell produces excess current. A promising candidate for the next generation space solar cells is a four-junction current matched metamorphic device with a Ge bottom cell [2]. In this structure, increased current from the Ge subcell would directly contribute to the power output of the multijunction cell. One feasible approach for this is decreasing the p-type doping level of the Ge substrate and improvement of the effective minority carrier lifetimes ($\tau_{eff}$) resulting in increased minority carrier diffusion length [3]. This increases the current generated by the longer wavelength photons. For this work, the decrease in p-type doping was realized by the Ge manufacturer Umicore and the improvement of $\tau_{eff}$ was realized by applying a surface passivation with the help of a $Si_xC_{1-x}$ layer stack [4,5]. The $Si_xC_{1-x}$ layer stack is designed in a way not only to increase $\tau_{eff}$, but also to act as a mirror layer for long wavelength photons, which induces two positive effects: (i) The light path and therefore the photon absorption between the direct and the indirect Ge bandgap (1550 nm and 1850 nm) is enhanced and therefore the generated current is further increased [6] and (ii) the mirror layer reflects photons with energies below the indirect Ge bandgap back into space leading to a solar cell temperature decrease and therefore higher open circuit voltage and efficiency [7].

One important issue for space solar cells is the End-Of-Life (EOL) performance as the satellites need a reliable power source for the controlled reentry at the end of their lifetime. Due to the earth's magnetic field, the Van Allen belts trap energetic electrons and protons, which are responsible for the overall degradation in space solar cell efficiency [8]. If a charged particle passes through a material, it transfers energy to the crystal lattice by ionization or atomic displacements. While energy loss by ionization dominates the particle range (penetration depth) in a material, the displacement is more critical for the performance of a solar cell, as the introduction of defects (vacancies, interstitials, etc.) gives rise to recombination centers that degrade the minority carrier lifetime and therefore the diffusion length.

Irradiation effects in TJ GaInP/GaInAs/Ge cells have been well studied at room temperature (RT) [9–11] or at low temperature [12,13]. In these studies, the Ge doping level is around $10^{18}$ at.cm$^{-3}$. At this high doping level, lifetime and mobility of minority carriers in Ge are small [14]. Irradiation will not induce drastic changes of Ge properties, because the number of defects produced during irradiation ($10^{14}$ - $10^{15}$ cm$^{-3}$) can be considered as a perturbation compared to the amount of doping atoms. In contrast to that, if the diffusion length is strongly increased by decreasing the Ge p-doping level down to $10^{16}$ at.cm$^{-3}$, the number of induced defects cannot be considered negligible any longer. Therefore, the radiative environment in space could become an issue for the development of such new Ge bottom cell technologies. The nature and content of electrically active defects induced by irradiation could drastically change the effective lifetime and diffusion length of minority carriers inside lowly p-doped Ge samples. In addition, the knowledge of point defects in Ge bulk is still far behind Si [15]. The reason is that optical and magnetic spectroscopy techniques (electron paramagnetic resonance (EPR), luminescence, IR absorption, …) that allowed to identify most of the simple and complex defects in Si, are not suitable for Ge. EPR lines in Ge are too weak due to short spin-lattice relaxation time, and too broad because of hyperfine and super hyperfine interactions with the different isotopes of Ge [16,17]. IR absorption bands are difficult to observe because of the lack of vibrational modes associated with oxygen-related defects [18–20]. As a consequence, most information on radiation induced defects in Ge has been obtained from deep level transient spectroscopy (DLTS) measurements [21–23].

The goal of this work is to analyze and understand the influence of irradiation on surface passivated p-doped samples. For that purpose, we performed room temperature irradiation using 1 MeV electron and proton accelerated ageing at fluences corresponding to EOL conditions in space for such devices. An effective minority carrier lifetime response of the Ge bulk and its surface is measured before and after irradiation using microwave photo conductance decay (μW-PCD) mappings in order to observe the changes as a function of the nature of incident particles, fluences, Ge p-doping levels, Ge wafer thickness and passivation treatments.

## 2 Experiments

Dislocation free Czochralski-grown 4" **Ge wafers** with different doping levels were provided by the manufacturer Umicore. The wafer specifications are summarized in Tab. 1. For a 1st electron and a 1st proton irradiation campaign, three doping levels were investigated. In a 2nd electron irradiation campaign, only lowly doped samples with three different wafer thicknesses were irradiated.

The deposition of the **$Si_xC_{1-x}$ layer stack for surface passivation** will be only summarized here, more details can be found in previous works [4,5]. The native oxide layer on the Ge surface was removed by a dry etching step under vacuum conditions inside a plasma enhanced chemical vapor deposition (PECVD) reactor using a $H_2$/Ar gas mixture. The amorphous $Si_xC_{1-x}$:H layer ($x \approx 0.95$) was deposited on both wafer surfaces using methane ($CH_4$), silane ($SiH_4$), hydrogen ($H_2$) and argon (Ar) as precursor gases at 270°C. A second stoichiometric a-SiC:H layer ("mirror" layer) was deposited using the same precursors. Stoichiometric SiC was chosen as mirror layer as it is known to be radiation hard [24]. The structure of such a lifetime sample is depicted in Figure 1.

Table 1: Ge wafer specifications used for this work in the 1st and 2nd irradiation campaigns.

| Name | Doping / at.cm$^{-3}$ | Thickness / μm | Mobility $\mu$ / cm².V$^{-1}$s$^{-1}$ [25] | Variation mirror layer? |
|---|---|---|---|---|
| 1st Irradiation campaign (electrons and protons) | | | | |
| Low | ~2.3x10$^{16}$ | 500 | 3120 | electrons: yes protons: no |
| Medium | 5.2x10$^{16}$-6.2x10$^{16}$ | 650 | 2440 | electrons: yes protons: no |
| High | 9.7x10$^{16}$-1.1x10$^{17}$ | 650 | 1860 | electrons: yes protons: no |
| 2nd Irradiation campaign (electrons) | | | | |
| Very low | ~1x10$^{16}$ | 500/300/150 | 3470 | no |

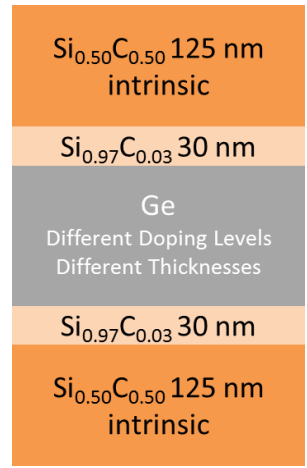

Figure 1: Structure of Ge lifetime samples for all irradiations with passivation and mirror layers on both sides. Ge wafers with different doping levels and thicknesses were used.

The electrical quality of the surface passivation procedure was characterized using the **microwave-detected photoconductance decay (µW-PCD)** technique [26], which is a purely transient technique measuring the exponential decay of excess minority carriers immediately after a short laser illumination. The used *Semilab* setup applies laser pulses at a wavelength of 904 nm. We estimated the excess carrier density after the laser pulse (injection level) to be $2 \times 10^{16}$ cm$^{-3}$. For the Ge samples in this study, we observed sensible measured distributions and reasonable scatter down to a measured lifetime level of about ~ 5 µs.

For the 1 MeV electron and proton irradiation campaigns, more than 300 samples with $2 \times 2$ cm$^2$ area were cut out of the 4" Ge wafers by chip sawing (DISCO dicing saw).

Both **electron irradiation campaigns** were conducted with a 1 MeV electron flux of $5 \times 10^{11}$ cm$^{-2}$.s$^{-1}$ at the SIRIUS irradiation facility (Palaiseau, France). The beam size at the surface of the sample is a 26 mm diameter disk measured on a microscope glass lamella using a circular 23 mm diaphragm before the window of the irradiation chamber. During each irradiation, currents reaching the sample and the diaphragm were recorded. The electron charge at the back of the sample was integrated using an Ortec 439 digital current integrator. The sample temperature during irradiation was kept at around 17°C by means of water-cooling. About 200 samples with an area of 4 cm² were irradiated with fluences between $3 \times 10^{13}$, and $1 \times 10^{16}$ e.cm$^{-2}$. The chosen values for electron energy, flux and fluences are typical for accelerated ageing simulations of space solar cells. In order to check the influence of the sample mounting, one sample per batch was only mounted on the sample holder without irradiation. The µW-PCD mappings of these samples will appear on the following µW-PCD mappings with the label "mounted".

For the **1st electron irradiation campaign**, lifetime samples with three different Ge doping levels (cf. Tab.1) were irradiated with fluences between $3 \times 10^{13}$ and $1 \times 10^{16}$ e.cm$^{-2}$. Irradiations were done in several iterations in order to get reliable results as summarized in Tab. 2.

Table 2: Number of Ge lifetime samples of three doping levels irradiated at different electron fluences during 1st electron irradiation campaign.

| 1st 1 MeV Electron irradiation | | Total fluence (e.cm$^{-2}$) | | | | | |
|---|---|---|---|---|---|---|---|
| | | **$3.10^{13}$** | **$1.10^{14}$** | **$3.10^{14}$** | **$1.10^{15}$** | **$3.10^{15}$** | **$1.10^{16}$** |
| Doping | **Low** | 11 | 8 | 13 | 4 | 4 | 1 |
| | **Medium** | 10 | 8 | 12 | 4 | 4 | 1 |
| | **High** | 12 | 11 | 15 | 4 | 4 | 1 |

Additionally, the properties of the mirror layer were changed for the 1st electron irradiation campaign in order to investigate if the radiation damage can be influenced by the properties of this layer. In the first sample batch, the mirror layer is about 125 nm thick and intrinsic (R samples, see Figure 1 and Table 3). In the second sample batch, the mirror layer is about 200 nm thick and intrinsic (T samples). As for further solar cell processing p-doping of the Si$_x$C$_{1-x}$ layer stack is essential, pure H$_2$ can be substituted by diborane (B$_2$H$_6$) diluted in H$_2$ with 0.25% in order to dope the layers. This was done with the mirror layer in the third sample batch (D samples). The passivated samples of batch 1-3 were exposed to an annealing step at 400°C for 5 min on a hotplate in open air at ambient conditions. The applied budget is representative for further solar cell processes like metal/Ge contact annealing. The annealing step was extended to 30 min at 400°C for sample batch four (A samples).

Table 3: Variation of mirror layer properties for the 4 batches of lifetime samples for 1st electron irradiation campaign.

| Mirror layer variation | | Mirror layer (Si$_{0.50}$C$_{0.50}$) thickness | Mirror layer (Si$_{0.50}$C$_{0.50}$) doping | Anneal @400°C |
|---|---|---|---|---|
| Sample batch | "R" (Reference) | 125 nm | intrinsic | 5 min |
| | "T" (Thickness) | **200 nm** | intrinsic | 5 min |
| | "D" (Doping) | 125 nm | **p-type** | 5 min |
| | "A" (Anneal) | 125 nm | intrinsic | **30 min** |

For the **2st electron irradiation campaign,** very lowly doped Ge wafers ($1 \times 10^{16}$ at.cm³) with three different wafer thicknesses (150, 300, 500 µm) were irradiated with three different fluences ($1 \times 10^{14}$, $3 \times 10^{14}$, $1 \times 10^{15}$ e.cm$^{-2}$). Irradiations were done in several iterations in order to get reliable results as summarized in Tab. 4.

Table 4: Number of Ge lifetime samples of different doping levels irradiated at different electron fluences during 2nd irradiation campaign.

| 2nd 1 MeV Electron irradiation | | Total fluence (e.cm$^{-2}$) | | |
|---|---|---|---|---|
| | | **1.10$^{14}$** | **3.10$^{14}$** | **1.10$^{15}$** |
| Wafer thickness | **150 µm** | 6 | 6 | 6 |
| | **300 µm** | 6 | 6 | 6 |
| | **500 µm** | 6 | 6 | 6 |

Only one 1 MeV **proton irradiation campaign** was conducted at the JANNUS Orsay irradiation facility (CSNSM, Orsay, France). Irradiations were done in several iterations in order to get reliable results as summarized in Table 5. The particles flux was constantly monitored during the irradiation process. During the proton irradiation campaign - analogously to the 1st electron irradiation campaign - lifetime samples with three different Ge doping levels (cf. Tab. 1) were irradiated with three different fluences of 1.33x10$^{10}$, 4x10$^{10}$ and 1.33x10$^{11}$ p.cm$^{-2}$. Such proton fluence values are based on the above mentioned 3x10$^{13}$, 1x10$^{14}$, and 3x10$^{14}$ e.cm$^{-2}$ electron fluences. They were deduced from screened relativistic NIEL calculations [27], showing that 1 MeV proton can transfer about 3000 times more energy than 1 MeV electron in Ge, considering a threshold displacement energy of 15 eV [28,29].

Table 5: Number of Ge lifetime samples of different doping levels irradiated at different protons fluences.

| 1 MeV Proton irradiation | | Total fluence (p.cm$^{-2}$) | | |
|---|---|---|---|---|
| | | **1.33x10$^{10}$** | **4x10$^{10}$** | **1.33x10$^{11}$** |
| Doping | **Low** | 12 | 11 | 11 |
| | **Medium** | 12 | 12 | 11 |
| | **High** | 12 | 12 | 11 |

## 3 Results

### 3.1 Lifetime degradation under 1-MeV electron irradiation

In Fig. 2, a typical lifetime mapping of an array of 25x4 cm² lowly doped Ge samples is depicted. In Fig. 2 a), the sample names are indicated. The capital letter stands for the treatment of the mirror layer (cf. Table 3), while the numbers refer to the original position of the sample on the Ge wafer. Umicore specifies a value of $\tau_{eff} \approx 50$ µs for the lowly doped, unpassivated Ge wafers. The mapping shows that the Si$_x$C$_{1-x}$ passivation leads to a massive increase of $\tau_{eff}$ up to 200…300 µs. The mapping also shows that the lifetime values on every sample show a certain distribution. This is due to the fact that the passivation is strongly depending on the exact condition of the unpassivated Ge wafer surface. Any scratch from sample handling or any inhomogeneity of the oxide before the etching step will cause an inhomogeneity of the lifetime distribution. Note that the samples with a doped mirror layer (labeled with D2_*) show higher lifetimes than all the other mirror layer treatments. As this result was not reproducible in additional experiments, we attribute the higher lifetimes of the D2_* samples to simple process fluctuations and no further conclusion has been derived from it.

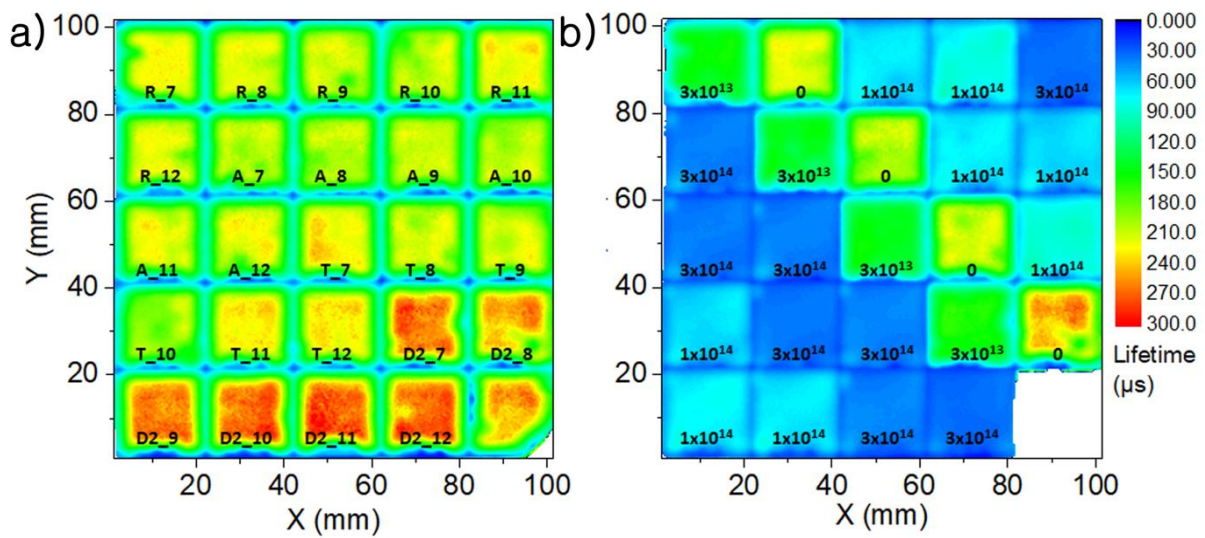

Figure 2: µW-PCD measurements mapping a) before and b) after electron irradiation for lowly doped Ge samples. In mapping a) the sample names and in b) the fluences in e.cm$^{-2}$ are indicated.

After irradiation, the samples and the reference samples were measured in the same arrangement and with the same measurement parameters as before irradiation and the mapping is depicted in Fig. 2 b). This method is appropriate to get a general overview of the degradation effect. In order to get more reliable lifetime results, the µW-PCD parameters were adapted to different lifetime ranges. The resulting data sets are used for the quantitative analysis reported on later. In Fig. 2 b), the irradiation fluences for every sample are indicated in the mapping. First of all, we state that the reference samples, which were shipped to the irradiation facility, back to ISE and measured again (labeled with an "0" in Fig. 2 b)), show exactly the same lifetime distribution as before irradiation. This means that both passivation and measurement setup were stable over time and shipping itself has no significant effect on lifetime values. 1 MeV electron irradiation induces a decrease in lifetime with increasing fluence between $3 \times 10^{13}$ and $3 \times 10^{14}$ cm$^{-2}$. For the highest fluence considered in this work ($1 \times 10^{16}$ e.cm$^{-2}$), the µW-PCD mappings (data not presented) show lifetime values ≤ 5 µs for all three Ge p-doping levels.

The decrease in lifetime with irradiation fluence is clearly observed by the µW-PCD mappings. To evaluate the lifetime mappings quantitatively, we proceeded as follows: First, all lifetime maps (before and after irradiation for three Ge doping levels and two electron irradiation campaigns) were separated in smaller maps of single samples of 2x2 cm² squares. Second, the lifetime values on the edges of these squares were discarded to account for edge recombination and potential artefacts. The width of the excluded "edge stripe" was chosen to be 2.5 mm for all samples. Third, the average and the standard deviation for all samples were calculated to evaluate the absolute lifetime before and after irradiation for each type of sample. For the average and the standard deviation, all lifetime values of the same sample type (doping, fluence,…) were used to obtain a good statistic. The evaluation shows that the properties of the mirror layer (thickness, doping, annealing) did not influence the degradation effect due to irradiation at all (R, T, D, A samples). Therefore, all R, T, D and A samples with the same doping level and irradiated at the same fluence were treated as a set of samples. The results can be found in Fig. 3.

The results in Fig. 3 show that $\tau_{eff}$ is lower for higher Ge p-type doping level. This is an expected result due to the influence of fermi level position on recombination.

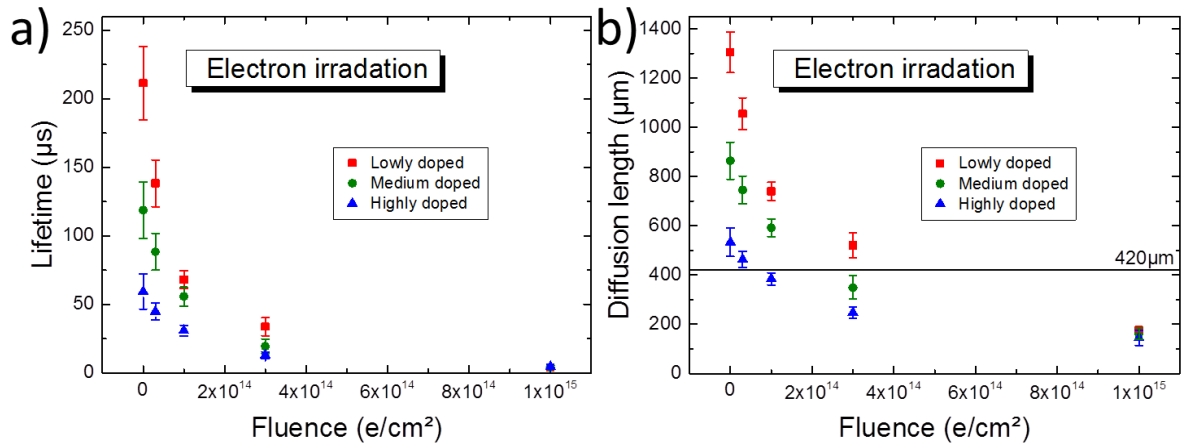

Figure 3: a) Evolution of effective lifetime in the Ge samples with three different doping levels (low, medium, high) under 1 MeV electron irradiations as a function of fluence and b) corresponding defective diffusion length.

The electron fluence shows a strong influence on the lifetime values of all Ge samples. At a fluence of 3.0x10$^{14}$ e.cm$^{-2}$, we observe a lifetime decrease by 85% in the irradiated samples for the lowest doping level. For fluences ≥ 1.0x10$^{15}$ e.cm$^{-2}$, $\tau_{eff}$ decreases to about 5 µs and thus reaches the detection limit of the µW-PCD method.

To assess the possible performance of the passivated, irradiated Ge as a future solar cell device, the effective diffusion length $L_D^{eff}$ can be derived from $\tau_{eff}$. $L_D^{eff}$ indicates the average distance that the minority carriers can travel before recombination. In Fig. 3 b), $L_\mathrm{D}^\mathrm{eff}$ is depicted as a function of doping level and fluence. $L_\mathrm{D}^\mathrm{eff}$ is calculated from $\tau_\mathrm{eff}$ with the following equation:

$$L_\mathrm{D}^\mathrm{eff} = \sqrt{\tau_{eff} * D}, \qquad (1)$$

$$D = \mu * \frac{k_B T}{e}. \qquad (2),$$

with *D* being the diffusion constant, $k_\mathrm{B}$ the Boltzmann constant, *T* the temperature and *e* the electron charge. The mobility *μ* of the minority carriers depends on the doping level $N_\mathrm{A}$. The mobility values in literature [25,30,31] result from measurements of Ge with different doping levels and show significant scatter that could be related to the crystallographic quality of the investigated wafers. For this work, we use the *μ*($N_\mathrm{A}$) values implemented in the solar cell calculation software PC1D [25] and the three doping levels listed in Tab. 1. With these mobility values, $L_\mathrm{D}^\mathrm{eff}$ was calculated according to Eq. (1) and (2) from the measured lifetimes $\tau_\mathrm{eff}$. This calculation assumes a homogeneous lifetime distribution throughout the samples, i.e. negligible surface recombination (compare section 3.3) and a weak depth dependence of the lattice damage caused by electron irradiation. Analogously to the lifetime, the estimated $L_\mathrm{D}^\mathrm{eff}$ depends on the doping level. Before irradiation, the lowly doped Ge wafers show a $L_\mathrm{D}^\mathrm{eff}$ of 1300 μm and for the highly doped wafers, $L_\mathrm{D}^\mathrm{eff}$ is still more than 500 μm. The reason for this long diffusion length is the good surface passivation quality and the high mobility in Ge. In a typical space solar cell, the Ge wafer is about 140 μm thick. As a rule of thumb, the diffusion length should be three times larger than the thickness of the wafer to allow for high quantum efficiency (420 μm). This value is indicated by a horizontal line in Fig. 3 b). It illustrates that an electron irradiation with a fluence of 1x10$^{14}$ e.cm$^{-2}$ is expected to cause no severe quantum efficiency losses in the Ge cell and even irradiation with 3x10$^{14}$ e.cm$^{-2}$ is expected not to induce efficiency losses for lowly doped Ge.

## 3.2   *μW-PCD lifetime degradation under 1-MeV proton irradiation*

To estimate the damage induced by 1 MeV protons within the SiC layer stack and the Ge bulk, we used the SRIM Monte Carlo program to simulate a monoenergetic, unidirectional beam of protons normally incident upon the surface of the samples [32]. Unlike 1 MeV electrons that produce sparsely distributed point defects in the target volume, 1 MeV protons induce displacement damage culminating in a peak (referred to as the Bragg peak) in the defect concentration at the end of the ion track.

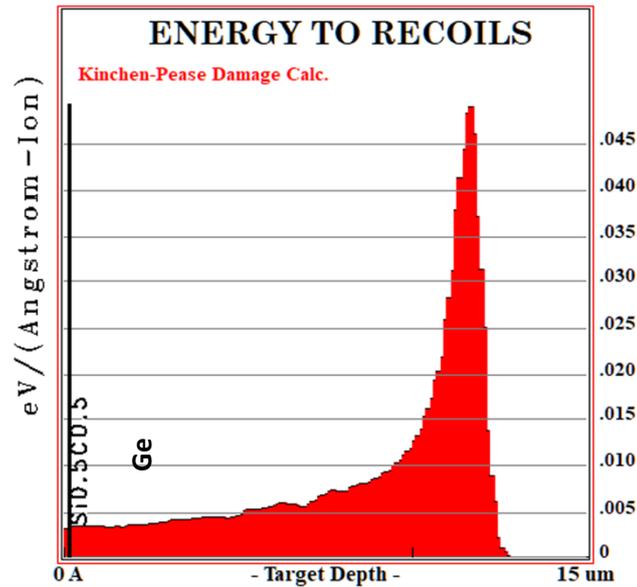

Figure 4: Rate of energy absorbed by recoil throughout the a-SiC layer stack and the Ge bulk by 1 MeV protons as calculated by SRIM [32].

The estimated penetration depth for 1 MeV protons is 12 µm. Fig. 4 presents the results of SRIM calculations which represent the case where back side protons stop in the Ge bulk penetrating the optical a-SiC layer. The "energy to recoil" data in Fig. 4 give the rate at which energy is absorbed by recoils that are produced per incident proton along the proton track through the materials. This data can be manipulated in order to deliver vacancy production and NIEL information. Integrating with respect to depth into the sample gives the total energy absorbed by the recoil events, which can then be used to calculate the total displacement damage induced in the materials per incident ion. Near the a-SiC/Ge interface, the energy deposition is linear and the amount of degradation is similar up to around 10 µm. In that case, degradation under proton irradiation is supposed to be equivalent to the results previously obtained with 1 MeV electron irradiation. By contrast, when we consider 1 MeV protons with a displacement damage peak around 12 µm inside the samples, a maximal deposition of energy occurs in this depth. This leads to an inhomogeneous lifetime distribution over the depth of the Ge wafers. To ensure that $\tau_{eff}$ measured with µW-PCD is a meaningful value anyway, we proceeded as follows: The samples were proton irradiated from the same side as they were µW-PCD measured. The measurement is done with a laser wavelength of 904 nm, which has a penetration depth of 0.4 µm in Ge. Thus, most generated excess charge carriers diffusing into the wafer to reduce the concentration gradient will necessarily pass the damaged layer. The measured $\tau_{eff}$ should therefore be a meaningful quantity to assess the damage caused by proton irradiation. However, it cannot be directly related to a homogeneous bulk lifetime.

The influence of 1 MeV proton irradiation analyzed by µW-PCD measurements is presented in Fig. 5 where the µW-PCD maps of lowly doped Ge lifetime samples before (Fig. 5 a)) and after (Fig. 5 b)) irradiation are depicted.

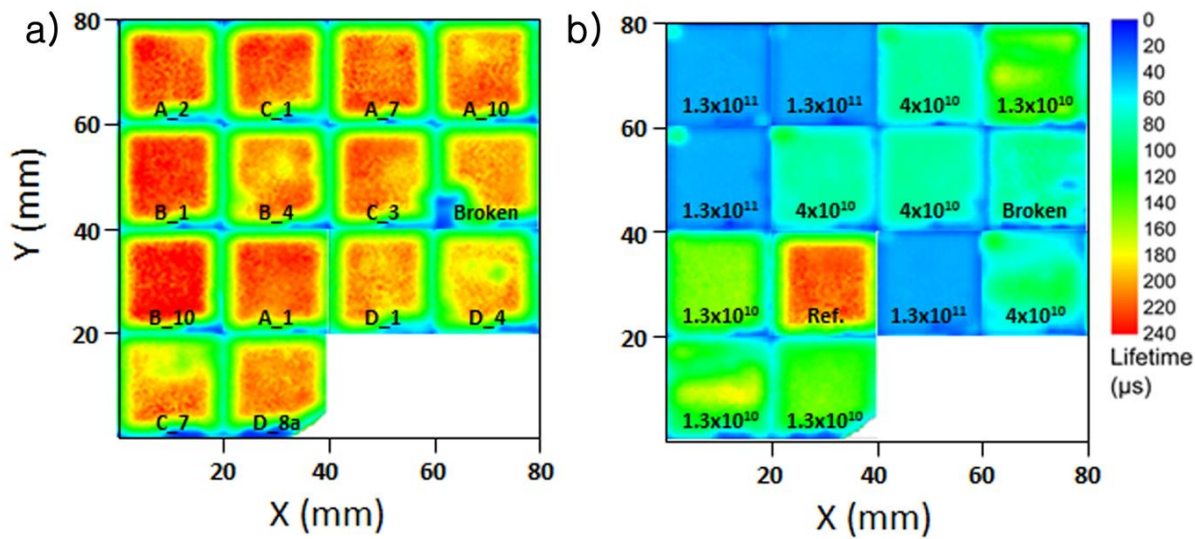

Figure 5: µW-PCD mappings of lowly doped Ge lifetime samples a) before and b) after 1-MeV proton irradiation. In a) the samples are labeled with the samples names and in b) the protons fluence is noted (in p.cm$^{-2}$).

For the proton irradiation samples, the passivation layer stack was not varied due to the result for electron irradiations which showed that the passivation layer treatment does not influence the degradation. The samples names (A_*, B_*, C_*, D_*) indicated in Fig. 5 a) refer to the different Ge wafers the samples come from. In Fig. 5 b), the samples are labeled with the proton fluences they were subjected to (in p.cm$^{-2}$). As previously observed with electron irradiation, the lowest proton fluence of 1.33x10$^{10}$ p.cm$^{-2}$ is sufficient to induce a significant decrease in the lifetime of lowly doped Ge passivated samples. For the highest proton fluence of 1.33x10$^{11}$ p.cm$^{-2}$, lifetime values lower than 50 µs are detected.

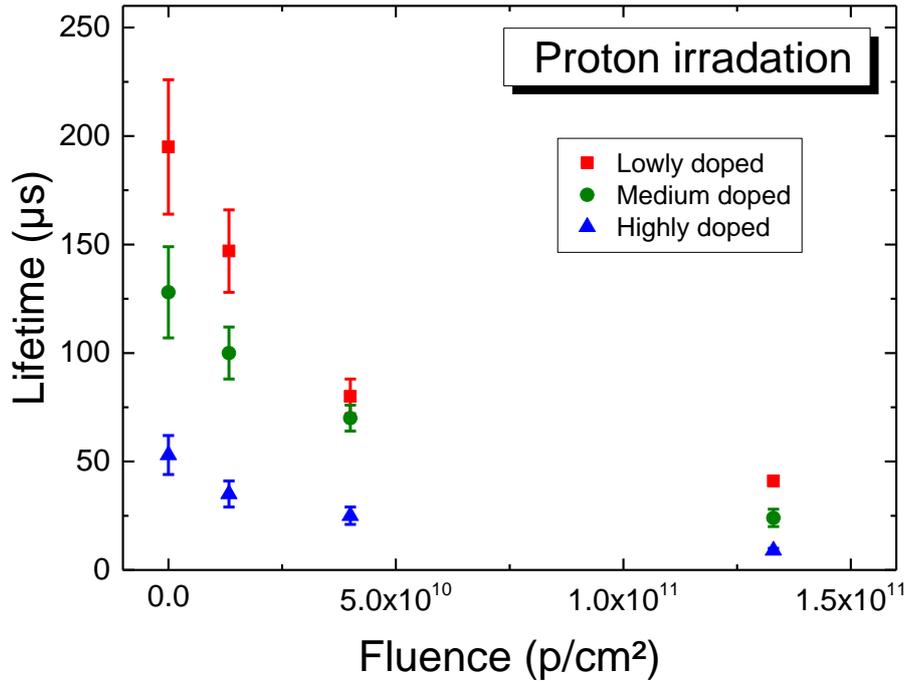

Figure 6: Evolution of lifetime in the Ge samples with three different doping levels (low, medium, high) under 1 MeV proton irradiation as a function of fluence.

Fig. 6 is a summary of all µW-PCD results on Ge lifetime samples before and after 1 MeV proton irradiation at different fluences and different p-doping levels. A clear correlation between the fluence and the lifetime can be observed as a function of the proton fluence. In contrast to electron irradiation (Fig. 3), the lifetime degradation after proton irradiation is less pronounced and no zero lifetime values are observed in the fluence range considered. As mentioned above, due to the inherently strong depth dependence of the lattice damage caused by protons, an evaluation of diffusion lengths $L_D^{eff}$ from $\tau_{eff}$ is less meaningful than for the above case of electron irradiation. However, our experiment clearly shows that the damage caused by proton irradiation of up to $1.33 \times 10^{11}$ p.cm$^{-2}$ does not lead to the recombination of all excess carriers. An irradiated Ge subcell would likely still be capable of current generation.

### 3.3  Influence of bulk and surface damage on lifetimes

In the 1$^{st}$ irradiation campaign, only the effective carrier lifetime $\tau_{eff}$ was determined, which combines the recombination in the bulk and at the surface of the Ge wafer. To distinguish between the influence of irradiation on bulk lifetime $\tau_{bulk}$ and surface carrier lifetime $\tau_{surface}$, a 2$^{nd}$ electron irradiation campaign was performed with Ge lifetime samples of different Ge wafer thicknesses $W$ (cf. Tab. 1). The p-type doping level of the Ge wafers irradiated in this campaign of $1 \times 10^{16}$ at.cm$^{-3}$, even lower than in the previous experiments. By determining the effective carrier lifetime $\tau_{eff}$ of these samples, a differentiation between $\tau_{bulk}$ and $\tau_{surface}$, characterized by the parameter surface combination velocity (*S*), is possible:

$$\frac{1}{\tau_{eff}} = \frac{1}{\tau_{bulk}} + \frac{1}{\tau_{surface}} = \frac{1}{\tau_{bulk}} + \frac{2}{W} \cdot S. \quad (3)$$

We chose Ge wafer thicknesses of 150 µm, 300 µm and 500 µm and plotted $\tau_{eff}$ versus 2/*W* to determine the carrier bulk lifetime $\tau_{bulk}$ as y-axis intersection and the surface recombination velocity *S* as slope of the applied fit.

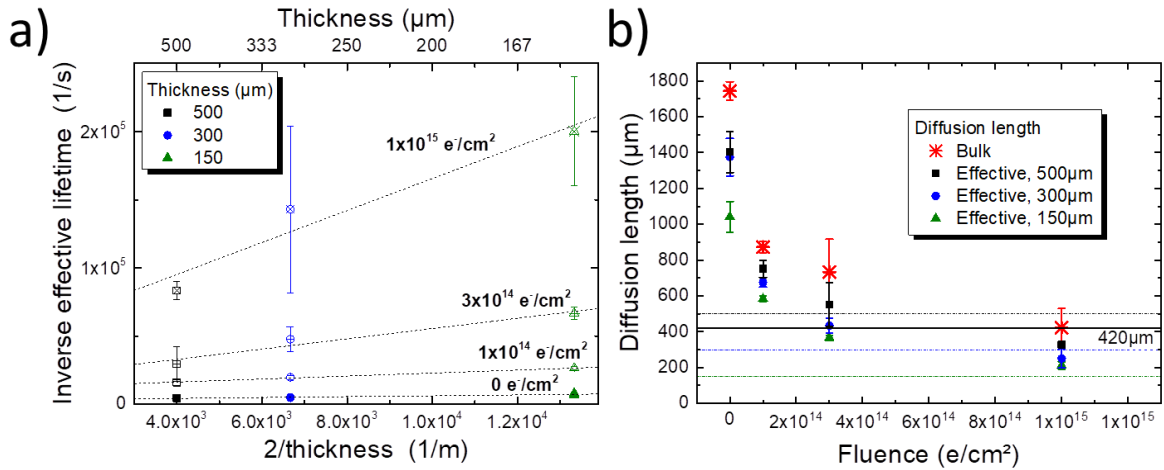

Figure 7: a) Determination of bulk lifetime $\tau_{bulk}$ and surface recombination velocity *S* for Ge lifetime samples before irradiation and after three different electron fluences. b) Effective diffusion length $L_D^{eff}$ derived from $\tau_{eff}$ and bulk diffusion length $L_D^{bulk}$ derived from $\tau_{bulk}$ as a function of the electron fluence. The dashed horizontal lines indicate the three Ge wafer thicknesses and the solid horizontal line indicates the three times the Ge wafer thickness in space solar cells.

Table 6: Values for $\tau_{bulk}$ and $S_{eff}$ after different electron irradiations determined from Fig. 7 a).

| Fluence (e.cm$^{-2}$) | *S* (cm.s$^{-1}$) | $\tau_{bulk}$ (µs) |
|---|---|---|
| 0 | 32 +/- 2 | 340 +/- 20 |
| 1 x 10$^{14}$ | 110 +/- 9 | 85 +/- 6 |
| 3 x 10$^{14}$ | 380 +/- 90 | 60 +/- 30 |
| 1 x 10$^{15}$ | 1200 +/- 300 | 20 +/- 10 |

This fit can be found in Fig. 7 a) for the samples before electron irradiation and after three different electron fluences. The corresponding values for $\tau_{bulk}$ and *S* are summarized in Tab. 6. The strong decrease for $\tau_{bulk}$ from more than 300 µs before irradiation down to about 20 µs after an electron irradiation with a fluence of 1x10$^{15}$ e.cm$^{-2}$ was expected. Surprisingly, *S* increases simultaneously from about 30 cm/s to more than 1000 cm.s$^{-1}$.

The effective diffusion length $L_D^{\text{eff}}$ was calculated from $\tau_{\text{eff}}$ for the three wafer thicknesses (500 µm, black squares, 300 µm blue circles, 150 µm green triangles) and is depicted in Fig. 7b). Additionally, the bulk diffusion length $L_D^{\text{bulk}}$ was calculated from $\tau_{\text{bulk}}$ (red stars) and added to Fig. 7b). $L_D^{\text{eff}}$ is higher (≈ 1400 µm BOL) than in the 1st irradiation campaigns, as the Ge doping is lower and therefore the mobility is higher (cf. Tab. 1). $L_D^{\text{bulk}}$ is even higher (≈ 1700 µm BOL), as the influence of recombination at the surfaces is eliminated in this case. $L_D$ decreases with increasing fluence and is smaller in thinner wafers, due to the bigger influence of the surfaces. The dashed horizontal lines in Fig. 7b) indicate the three Ge wafer thicknesses and the solid horizontal line indicates a thickness of 3 times 140 µm. Considering $L_D^{\text{bulk}}$, a complete current collection is expected even after the highest irradiation fluence of 1x10$^{15}$ e.cm$^{-2}$ as the bulk diffusion length still exceeds three times the Ge wafer thickness in a solar cell (140 µm). Considering $L_D^{\text{eff}}$, an irradiation with 1x10$^{14}$ e.cm$^{-2}$ is expected to cause no significant losses in the Ge cell. At 3x10$^{14}$ e.cm$^{-2}$, a complete current collection is still expected for the 500 µm and 300 µm thick wafer. For 150 µm, losses will occur but are still expected to be minor. In the discussion section, we will address the question if $L_D^{\text{eff}}$ or $L_D^{\text{bulk}}$ is the more relevant parameter for the space solar cell.

## 4  Discussion

### 4.1  *Lifetime degradation in Ge due to electron/proton irradiation*

Ge wafers with Si$_x$C$_{1-x}$ surface passivation show high minority carrier lifetimes before electron (Fig. 3) and proton (Fig. 6) irradiation. The Beginning Of Life (BOL) lifetime depends on the doping level of the Ge wafers, due to increasing intrinsic recombination. Typical BOL lifetime values in this work are ≈ 200 µs for lowly doped Ge (2x10$^{16}$ at.cm$^{-3}$), ≈ 120 µs for medium doped Ge (5x10$^{16}$ at.cm$^{-3}$) and ≈ 50 µs for highly doped Ge (1x10$^{17}$ at.cm$^{-3}$). To the authors' knowledge this is the highest reported BOL lifetimes so far for a Ge irradiation study. This high starting lifetime allows for the detection of bulk lifetime decreases during both electron and proton irradiation for all three Ge doping levels and for the evaluation of different types of Si$_x$C$_{1-x}$ passivation stacks. After an electron fluence of 1x10$^{15}$ e.cm$^{-2}$, the degradation extent surpasses the detection limit of the µW-PCD tool (Fig. 3). At this fluence, the lifetimes for all three doping levels are thus likely below 5 µs. The highest proton fluence used in this work (1.3x10$^{11}$ p.cm$^{-2}$) did not result in such strong degradation and the dependence of lifetime on doping level is still detectable (Fig. 6). However, the trend shows an unambiguous lifetime decay for the three Ge doping levels as well with the measured lifetimes converging with increasing fluence.

The decrease of lifetime due to irradiation can be explained as follows: One 1 MeV electron will on average displace one Ge atom from its lattice position generating a vacancy and an interstitial atom - a so-called Frenkel pair. Such defects can act as recombination centre for electron-hole pairs and thereby decrease the minority carrier lifetime [33]. 1 MeV protons also cause displacement of Ge atoms from their lattice position. However, the displaced Ge atom usually has enough energy to cause further displacements in the lattice resulting in so-called cluster damage. One 1 MeV protons deposit 3 orders of magnitude more energy in Ge than 1 MeV electrons [28,29]. While the point defects induced by electron irradiation are distributed homogeneously throughout the whole Ge bulk, the point defects induced by protons are mostly located about 12 μm under the passivated surface (c.f. Fig. 4) due to the different interaction cross sections. The choice of the electron and proton fluences was made in a way to ensure a similar irradiation defect production for both, electrons and protons. However, a real calculation for equivalent fluences can only be made in the linear regime of energy deposition for both electron and proton irradiation. This is the case for III-V solar cells with a thickness < 10 μm but not for Ge bulk with a thickness of ≥ 500 μm. Therefore, no direct assignment and comparison between electron and proton fluences is possible. That means that we can only state that the general degradation behaviour due to the homogenous defect production of electrons and the localized defect production of protons in the Ge bulk is similar. However, we cannot answer the question if there any differences in detail.

Another difference of electron and proton irradiation is the fact that the protons remain in the Ge bulk as H atoms. Therefore, formation of H related defects or interactions of H with present defects are expected to follow proton irradiation. It is beyond the scope of this work to elucidate the impact of the spatial defect distribution and the presence of H atoms on the observed degradation further.

During the 2$^{nd}$ electron irradiation campaign, we observed that the irradiation does not only reduce the bulk lifetime, but also increases the surface recombination velocity *S*. This means that irradiation does not only lead to Ge bulk defects, but also to defects at the a-Si$_x$C$_{1-x}$:H/Ge interface. An in-depth characterization of the interface is beyond the scope of this work. However, irradiation introduced breakage of Ge-Si and Ge-H and resulting dangling bonds are the most evident process at the a-Si$_x$C$_{1-x}$:H/Ge interface.

### *4.2 Correlations between lifetime and irradiation defects*

As described in the introduction, the identification of defects in Ge with common methods is particularly difficult. We therefore compare our findings to similar studies in literature to find out about the nature of the irradiation defects. First of all, according to older literature, the introduction of irradiation defects into p-type doped Ge at room temperature was not expected [17,19,33–35]. This is due to the fact that for the "classical" methods like EPR, DLTS and IR spectroscopy, some kind of defects in Ge are invisible [36,37]. Newer methods like XRD [36], DLTS at n⁺p Ge diodes [21] or carrier lifetime measurements of passivated Ge samples [14] show unambiguously that irradiation induced defects are present. The present knowledge about irradiation introduced point defects in Ge is summarized in the review paper from Emtsev *et al.* [37].

The nature of the defects in irradiated p-type Ge is still under discussion. It is possible that a minor amount of Frenkel pairs, which are doubtlessly the primary defect after irradiation, remains present even at room temperature [36]. Another possibility is the formation of room temperature stable di-vacancies, as vacancies in Ge are very mobile at room temperature [19,21].

While we cannot say more about the nature of the created defects at present, we want to point out that this study demonstrates lifetime measurements of passivated Ge samples to be a simple and sensitive method to detect defects. While the atomic nature or the energetic level of the created defects are not accessible, lifetime measurements can be directly related to solar cell performance. Furthermore, more elaborate measurement and evaluation of charge carrier lifetimes can give access to defect parameters as exercised in abundance in silicon-based technology [38,39].

### *4.3  Influence of lifetime on solar cell performance*

We have seen that the nature of defects in Ge after irradiation is still unclear. However, the more important question for the application of the surface passivation in space solar cells is, for which fluences the passivation is still good enough to collect all the current from the Ge subcell.

In the case of electrons irradiated samples, this can be answered with the help of the diffusion length $L_\mathrm{D}$, which can be derived from the minority carrier lifetime $\tau$. As for our lifetime samples the effective lifetime $\tau_\mathrm{eff}$ is measured, the effective diffusion length $L_\mathrm{D}^\mathrm{eff}$ is derived which combines the influence of the bulk and the surface recombination. $L_\mathrm{D}^\mathrm{eff}$ is easy to access and a good estimation for diffusion length in the final solar cell. However, the more relevant value for the solar cell performance is the bulk diffusion length $L_\mathrm{D}^\mathrm{bulk}$, which only depends on the recombination in the Ge bulk and not on the recombination at the surfaces. This is comparable to the situation in a solar cell, where a carrier reaching the rear side of the cell will most likely be collected by the metal contacts. However, the access of $L_\mathrm{D}^\mathrm{bulk}$ is more complex, as it requires $\tau_\mathrm{bulk}$ and therefore a thickness variation of lifetime samples. Therefore, $L_\mathrm{D}^\mathrm{eff}$ was calculated for all electron irradiated samples in this work, whereas $L_\mathrm{D}^\mathrm{bulk}$ was only derived for the samples of the 2$^\mathrm{nd}$ electron irradiation (doping of 1x10$^{16}$ at.cm$^{-3}$). $L_\mathrm{D}^\mathrm{eff}$ is depicted in Fig. 3b and 7b) and $L_\mathrm{D}^\mathrm{bulk}$ is depicted in Fig. 7b). The effective diffusion length is higher for lower doping concentrations, decreases with increasing fluence and is smaller in thinner wafers, due to the increasing influence of the surfaces. When the influence of the surface is corrected for ($L_\mathrm{D}^\mathrm{bulk}$), the diffusion length is found to depend on the fluence and become independent of the wafer thickness.

The solid horizontal lines in Fig. 3b) and 7b) indicate a diffusion length of three times the cell thickness of 140 µm, which corresponds to the $L_\text{D}$ value required for good solar cell performance. The determination of $L_\text{D}^\text{bulk}$ for wafers with a doping level of 1x10$^{16}$ at.cm$^{-3}$ predicts no significant current losses even for the highest electron fluence of 1x10$^{15}$ e.cm$^{-2}$. However, even for the less optimistic parameter $L_\text{D}^\text{eff}$ we observe that for all doping levels and wafer thicknesses, an irradiation with 1x10$^{14}$ e.cm$^{-2}$ is expected to cause negligible quantum efficiency losses in the Ge cell. For 3x10$^{14}$ e.cm$^{-2}$, a good current collection can still be expected for the 500 µm and 300 µm lowly doped wafers, while the higher doped material would suffer losses.

For 1 MeV proton irradiation, the assessment of $L_\text{D}^\text{eff}$ from $\tau_\text{eff}$ is not meaningful due to the strong inhomogeneity along the Ge wafer depth. Therefore, for proton irradiation, no direct conclusion concerning the performance of the corresponding solar cell can be drawn. More detailed investigations on the lifetime distribution and solar cell simulations would be necessary but are beyond the scope of this work. In an upcoming study of some of the authors the impact of irradiation will be quantified by the irradiation of solar cells with lowly doped, passivated Ge wafer subcells and by corresponding solar cell simulations.

However, even after the highest proton fluence of 1.33x10$^{11}$ p.cm$^{-2}$ still considerable lifetimes were measured. Due to the high mobility in lowly doped Ge, we therefore expect still significant solar cell performance. However, this statement must be verified experimentally in the future.

We conclude for both, electron and proton irradiation, that despite the strong lifetime decrease, an improvement of the Ge cell current is expected for space solar cells with medium exposure time to irradiation, such as Low Earth Orbit Constellations and High Altitude Pseudo Satellites (HAPS).

## 5   Conclusion

In this work, we investigated bulk and surface properties of germanium for the use in next generation space solar cell devices. For that purpose, Ge substrates with a wide doping concentration range from 1x10$^{16}$ to 1x10$^{17}$ at.cm$^{-3}$ and a passivating SiC layer were irradiated using 1 MeV electrons and protons. Minority carrier lifetimes were measured by means of µW-PCD as a function of the fluence, the nature of the irradiation particle, different Ge doping levels and different Ge thicknesses. The investigated samples featured lifetimes ranging from ≈ 50 to 230 µs initially (BOL) down to few µs after irradiation (EOL). Although a direct comparison between electron and proton fluences is not meaningful, the trend in lifetime decrease is similar. This indicates clearly that a vast amount of defects was created in Ge irradiated at 17°C and stored at room temperature. Such defects were not reported in "classical" literature due to missing characterization techniques and missing surface passivation and just started to appear in literature in the last years.

From our experiments, we cannot draw conclusions about the nature of the defects and therefore assume in accordance to recent literature the presence of Frenkel pairs and divacancies in our samples. From a Ge thickness variation experiment, we concluded that the irradiation causes recombination centers in the Ge bulk as well as at the surface. The latter are most likely dangling bonds introduced by Ge-H/Ge-Si bond breakage.

For electron irradiated samples, the diffusion lengths derived from the lifetime measurement gives rise to the assumption that no major quantum efficiency decrease will take place in future space solar cells for electron irradiations up to $1 \times 10^{15}$ e.cm$^{-2}$. We cannot directly transfer the used approach to proton irradiation and thus cannot state rigorous conclusions on quantum efficiency. However, based on our lifetime measurements, we expect significant retained cell performance after proton irradiation up to $1.3 \times 10^{11}$ p.cm$^{-2}$. Based on these findings, we expect lowly doped, passivated Ge wafers to be a promising candidate for subcells in next generation space solar cells for applications where moderate irradiation is expected.

## Acknowledgment

This work has received funding from the European Union's Horizon 2020 research and innovation program within the project SiLaSpaCe under grant agreement No 687336. SIRIUS and JANNUS Orsay teams are thanked for the electron and proton irradiation campaigns.